\documentclass[10pt,oneside]{amsart}
\usepackage{adjustbox}

\usepackage{tikz}
\usetikzlibrary{automata, positioning, arrows}

      \usepackage{amssymb}
\usepackage{amsfonts,amstext}
\usepackage{graphicx}
\usepackage{url}
\usepackage{amsmath,amstext,amssymb,amscd}
\usepackage{verbatim} 
\usepackage{mathtools}
      \usepackage{tabularx,booktabs}
      \usepackage{multicol}

\usepackage{amssymb,latexsym}
\usepackage{amsmath}
\usepackage{amsthm}
\usepackage{amssymb}
\usepackage{makecell}
\newtheorem{thm}{Theorem}[section]

\theoremstyle{remark}

\theoremstyle{definition}

\newtheorem{ex}[thm]{Example}

\usepackage{amsthm}
\allowdisplaybreaks

\usepackage{multicol}

      \makeatletter
      \def\@setcopyright{}
      \def\serieslogo@{}
     
      \makeatother

\begin{document}

\author{David McCune}
\address{David McCune, Department of Mathematics and Data Science, William Jewell College, 500 College Hill, Liberty, MO, 64068-1896}
\email{mccuned@william.jewell.edu}

\title[Ranked Choice Bedlam in Oakland]{Ranked Choice Bedlam in a 2022 Oakland School Director Election}

\begin{abstract}
The November 2022 ranked choice election for District 4 School Director in Oakland, CA, was very interesting from the perspective of social choice theory. The election did not contain a Condorcet winner and exhibited downward and upward monotonicity paradoxes, for example. Furthermore, an error in the settings of the ranked choice tabulation software led to the wrong candidate being declared the winner. This article explores the strange features of this election and places it in the  broader context of ranked choice elections in the United States.
\end{abstract}

 \subjclass[2010]{Primary 91B10; Secondary 91B14}

 \keywords{ranked choice voting, monotonicity, Condorcet}

\maketitle

\section{Introduction}

On December 9, 2022, Nick Resnick assumed the office of the District 4 School Director in Oakland, CA, after having won a ranked choice election in November. On December 28, the Registrar of Voters in Alameda County (the county in which Oakland is located) announced that the ranked choice tabulation software responsible for counting votes had not been configured properly, resulting in incorrect vote counts for the three official candidates in the race. Once the tabulation setting was fixed, the software output a new winner: Mike Hutchinson. For the first documented time in the history of ranked choice voting (RCV) in the US, a ballot processing error caused the wrong candidate to be named the winner. Resnick resigned rather than engaging in a lengthy legal battle, and a judge allowed Mike Hutchinson to assume the office in March 2023.

Interestingly, even if the ballots had been processed correctly this election would still be one of the most fascinating ranked choice elections in US history, at least from the perspective of social choice theory. For example, the election did not contain a Condorcet winner and exhibited downward and upward monotonicity paradoxes, outcomes which are rarely observed in real-world ranked choice elections. The purpose of this article is to examine this election's strange features, while contextualizing them within the landscape of American ranked choice political elections.

\section{What is RCV?}

In an RCV election, voters submit ballots ranking the candidates in order of preference. The Oakland School Director election contained the three (not write-in) candidates Mike Hutchinson, Pecolia Manigo, and Nick Resnick, and voters cast preference ballots with a ranking of these candidates. For example, a voter could cast a ballot ranking Hutchinson as their first choice, Manigo as their second choice, and Resnick as their third choice. We use the notation $\succ$ to denote when a voter preferences one candidate to another, and thus such a ballot is denoted Hutchinson$\succ$Manigo$\succ$Resnick. Voters were not required to submit a complete preference ranking, and so could cast a ballot simply ranking Hutchinson as their first choice and leave the other rankings blank. 

\begin{table}[]
  \centering
\begin{tabular}{c|c|c|c}
\textbf{Num. Voters} & \textbf{Choice 1} & \textbf{Choice 2} & \textbf{Choice 3}\\
\hline
2283 & Hutchinson & Manigo & Resnick\\
1280& Hutchinson & Manigo& $-$\\
1807 & Hutchinson & Resnick & Manigo  \\
530 & Hutchinson & Resnick & $-$ \\
2327 & Hutchinson &$-$&$-$\\
1734&Manigo & Hutchinson & Resnick\\
2460&Manigo & Hutchinson &$-$\\
1421&Manigo&Resnick & Hutchinson\\
729&Manigo&Resnick &$-$\\
1846&Manigo & $-$&$-$\\
2171& Resnick&Hutchinson&Manigo\\
924& Resnick&Hutchinson&$-$\\
2246&Resnick&Manigo&Hutchinson\\
934&Resnick&Manigo&$-$\\
3740&Resnick & $-$ & $-$\\

\end{tabular}

\caption{The November 2022 District 4 Oakland School Director preference profile, neglecting write-in candidates. The $-$ symbol denotes a blank ranking.}
  \label{preference_profile}
\end{table}

In American ranked choice political elections vote data is shared via a \emph{cast vote record},  an anonymized record of the preference ranking provided by each voter. The cast vote record for the Oakland School Director election can be found at \cite{O2}. To display the data in a compact fashion we aggregate the ballots into  a \emph{preference profile}, which shows the number of each type of cast ballot. Table \ref{preference_profile} shows the preference profile of the Oakland School Director election, in which we've disregarded write-in candidates in order to keep the table compact. The number 2283 denotes that 2283 voters cast a ballot of the form Hutchinson$\succ$Manigo$\succ$Resnick; the other numbers convey similar information about the number of voters who cast the corresponding type of ballot.

To choose a winner of an election given a preference profile, RCV proceeds in a series of rounds. In each round, a candidate's first-place votes are counted; a candidate with a majority of the votes is declared the winner. If no candidate achieves  a majority then the candidate with the fewest first-place votes is eliminated and their votes are transferred to the next candidate on their ballots who has not previously been eliminated. The method continues in this fashion until a candidate achieves a majority of the remaining votes. To illustrate this process, note from Table \ref{preference_profile} that Hutchinson receives 8227 first-place votes (the sum of the first five numbers), Manigo receives 8190, and Resnick receives 10,015. Manigo receives the fewest and is therefore eliminated, and $1734+2460=4194$ of her votes are transferred to Hutchinson while $1421+729=2150$ are transferred to Resnick. After the vote transfers, Hutchinson defeats Resnick 12421 votes to 12165, and therefore Hutchinson should have been declared the winner  instead of Resnick.

\section{Ballot Processing and Getting the Wrong Winner}\label{sec:ballots}

Table \ref{preference_profile} displays the ballot data in a very clean fashion. In practice, it is not uncommon for voters to cast ballots which do not convey ranking information so cleanly. For example, suppose a voter casts the ballot Manigo$\succ-\succ$Hutchinson. How should we interpret such a ballot? Clearly Manigo is this voter's top choice, but the skipped ranking in the middle makes the voter's intentions unclear. Is this voter saying that Hutchinson is their least favorite, and so should we infer that Resnick is this voter's second choice? Or should we infer that the voter was attempting to cast the ballot Manigo$\succ$Hutchinson$\succ$ $-$, placing Hutchinson in second and leaving Resnick off the ballot? An elections office must decide how to handle such ballots, and there is not an obvious best way to do so. In this section we briefly discuss the two most common issues on ranked choice ballots: skipped rankings and a ranking on a ballot where a voter gives two candidates that same ranking (which is referred to as an \emph{overvote} in that ranking).

For our discussion, we consider a hypothetical election with five candidates $A$, $B$, $C$, $D$, and $E$.

\textbf{Skipped Rankings}. Suppose a voter casts the ballot $A\succ - \succ B\succ C \succ D$. In some sense this voter has made an error by skipping the second ranking, and we must decide how to handle this error. As far as I am aware, every elections office in the US which runs RCV elections interprets this ballot as $A\succ B\succ C \succ D\succ - $, simply ignoring the skipped ranking and shifting candidates up. This seems like a reasonable way to interpret the voter's intentions. If a voter skips more than one ranking, however, it becomes less clear how to interpret the voter's ballot. For example, if a voter casts the ballot $A\succ - \succ - \succ - \succ B$, is it reasonable to shift $B$ up to the second ranking? Perhaps this voter is confused about how the ballots will be processed, and is communicating that $B$ is their least favorite candidate. In this case, perhaps we should read the ballot as having $A$ in the first ranking and all subsequent rankings as blank, including the ranking with $B$ listed. This is how the Division of Elections office in Alaska handles such ballots, for example: if a voter leaves two consecutive rankings blank, every candidate listed after these skipped rankings is disregarded.

The Alameda County Office of Elections decides to ignore any skipped rankings, simply shifting candidates up. Thus, the ballot $A\succ - \succ - \succ - \succ B$ is read as having $B$ ranked second.

\textbf{Overvotes}. Suppose a voter casts the ballot $A \succ B \textbf{ and } C \succ D \succ - \succ -$, making an overvote error in the second ranking. How should this ballot be interpreted, given the overvote? Most elections offices interpret this ballot as only having $A$ in the first ranking, disregarding the candidates in the overvote and any candidates ranked after the overvote. There are exceptions to this interpretation: the elections office in Minneapolis, MN, for example, treats an overvote like a skipped ranking, and thus would interpret this ballot as having $A$ in the first ranking and $D$ in the second ranking.

The Alameda County Office of Elections disregards the candidates listed in the overvote and any candidates ranked on the ballot after the overvote. Thus, they read the ballot $A \succ B \textbf{ and } C \succ D \succ - \succ -$ as having only $A$ ranked in first place, ignoring candidate $D$. That is, even if $A$, $B$, and $C$ were all eliminated from the election, this ballot would not be transferred to $D$ (in Minneapolis, if $A$ were eliminated then this ballot would be transferred to $D$). Once the tabulation software reaches a ranking with an overvote, it discards the ballot from further consideration.

I also note that, as far as I am aware, all elections offices which process preference ballots ignore a candidate in a ranking if that candidate was ranked previously on the ballot. For example, the ballot $A\succ A\succ B \succ A\succ C$ would be processed as $A \succ B \succ C \succ - \succ -$. If a candidate appears multiple times on the ballot, such an error does not cause the ballot to be removed from the counting process at some point, the way an overvote error would.

 Table \ref{ballot_examples} illustrates how the Alameda County Office of Elections processes ballots. Each of the ballots in the table should be treated as equivalent by the RCV tabulation software, with Ballots 2-6 all being transformed into Ballot 1.
 
  \begin{table}[]
  \centering
\begin{tabular}{c|c|c|c|c|c}
\textbf{Ballot} & \textbf{Choice 1} & \textbf{Choice 2} & \textbf{Choice 3} & \textbf{Choice 4} & \textbf{Choice 5}\\
\hline
Ballot 1 & $A$ & $B$ & $-$ & $-$ & $-$\\
Ballot 2 &$-$ & $-$& $A$ & $B$ & $-$\\
Ballot 3 & $-$ & $A$ & $-$ & $-$ & $B$\\
Ballot 4 & $A$ & $B$ & $C$ and $D$ & $E$ & $-$\\
Ballot 5 & $-$ & $A$ & $-$ & $B$ & $C$ and $D$\\
Ballot 6 & $A$ & $A$ & $B$ & $-$ & $B$\\

\end{tabular}

\caption{Examples of ballots which are all supposed to be treated as $A\succ B \succ -\succ- \succ -$ by the RCV tabulation software in Alameda County.}
  \label{ballot_examples}
\end{table}
 
 Now that we understand how ballots should be processed, we can address the tabulation error which occurred in this election. Write-in candidates received only 269 first-place votes, and thus the first step in the RCV process is to eliminate the write-ins and transfer these votes to the next candidate ranked on the ballot. However, the software was set so that if a ballot was either blank or contained a write-in candidate in the first ranking, the ballot was not counted when tallying the vote totals for the three official candidates after disregarding write-ins. For example, after disregarding write-ins,  ballots of the form $-\succ$Hutchinson$\succ \dots$ or Write-in$\succ$Hutchinson$\succ\dots$ were not counted for Hutchinson in the first ``real'' round of counting when tallying votes for the three official candidates, even though according to Oakland's balllot processing rules these ballots should shift Hutchinson up to the first ranking. There were 235 ballots in which the first ranking listed a write-in candidate or was skipped, and which also ranked one of the three official candidates somewhere on the ballot; none of these ballots were counted when calculating vote totals for the three official candidates after eliminating write-ins\footnote{A statement from the Alameda County Registrar of Voters says, ``\emph{$[$The tabulation software$]$ should have been configured to advance ballots to the next ranking immediately when no candidate was selected for a particular round. This means that if no candidate was selected in the first round on the ballot, then the second-round ranking would count as the first-round ranking, the third-round ranking would count as the second round ranking, and so on. For the November 2022 General Election, the setting on the County’s equipment counted the RCV ballots in the manner in which the ballot was completed, meaning no vote was registered for those ballots in the first round of counting because those voters did not identify a valid candidate in a particular rank on the ballot.}'' I confirmed with an elections official that the error also involved write-in candidates, as I describe.}. As a result of this error, the vote initial vote totals were 8112, 8153, and 9954 for Hutchinson, Manigo, and Resnick, respectively, after eliminating write-ins. According to these vote totals, Hutchinson received the fewest first-place votes and was eliminated, and Resnick defeated Manigo head-to-head in the final round. This is how Resnick was wrongly crowned the winner of the election.

\begin{ex}
 
\textbf{Aside: A 2021 City Council Election in Portland, Maine.} The tabulation software in Alameda County had the incorrect setting chosen, and as a result did not properly transfer votes for write-ins after they were eliminated. However, even if the ballots are processed correctly according to the chosen settings, it is possible that the way we choose to process ballots can affect the winner. It seems like a disturbing possibility that if ballots are processed in the manner of the Alameda Elections Office then we get one winner, but if we process the same set of ballots in the manner of the Minneapolis elections office then we get a different winner. To see how this could play out, consider the 2021 city council ranked choice election in Portland, ME, for the at-large council seat. This election contained the two front-running candidates Brandon Mazer and Roberto Rodriguez, as well as two weaker candidates. Portland processes ballots in the manner of Alameda County and, for the first time in the history of ranked choice elections in the US, when the ballots were fed into the tabulation software the result was a tie. Both Mazer and Rodriguez earned 8529 votes in the final round. A hand recount subsequently declared Rodriguez the winner by a handful of votes, but several ballots remained in dispute and Rodriguez won only after Mazer decided not to continue pursuing legal action\footnote{See the linked article, accessed on March 9, 2023, for a description of this election. \url{https://www.pressherald.com/2021/11/10/recount-continues-in-portland-at-large-council-race/}}.

If Portland processed ballots in the manner of the Minneapolis elections office, skipping overvotes rather than discarding the rest of a ballot after encountering an overvote (and also ignoring all skipped rankings), then Mazer would have won with 8552 votes to Rodriguez's 8539 in the final round. In fact, if the ballots were processed in any  reasonable way other than how Portland does with respect to the handling of skipped rankings or overvotes, the tabulation software would have declared Mazer the winner\footnote{The cast vote record for this election is available upon request from me or the Portland City Clerk's office if the reader wants to verify these outcomes.}.  Of course, a hand recount could still declare Rodriguez the winner in these cases, but under different ballot processing rules it is impossible to anticipate the results of such a recount. Thus, it seems that the choice of how to process ballots can affect the winner of an RCV election in a very close race, even if the tabulation software is calibrated correctly.

\end{ex}

In the Oakland School Director election, we checked that Hutchinson is the RCV winner regardless of how we choose to handle skipped rankings and overvotes according to the rules of different elections offices in the US. Thus, the primary issue in this election was the erroneous non-transferral of ballots with a skipped first ranking or with a write-in candidate ranked first. We note that 225 ballots contained an overvote in some ranking and 304 ballots satisfied that a voter skipped a ranking and then listed a candidate in a later ranking. While these ballots represent a very small proportion of the total ballots cast, in extremely close elections such as what occurred in Oakland and Portland it is possible for the winner to be affected by the choice of how to handle such ballots.

\section{Condorcet Issues and the Spoiler Effect}\label{sec:Cond}

Which candidate ``should'' win the Oakland School Director election, assuming ballots are processed correctly? Which candidate is the ``most deserving,'' given the ballot data in Table \ref{preference_profile}? Even though Hutchinson is the RCV winner, we could argue using head-to-head comparisons of candidates that Manigo should win instead. Note from Table \ref{preference_profile} that \[1734+2460+1421+729+1846+2246+934 = 11370\] voters prefer Manigo to Hutchinson while \[2283+1280+1807+530+2327+2171+924=11322 \] voters prefer Hutchinson to Manigo. (We assume that if a voter ranks only one candidate, the voter is indifferent among the candidates left off the ballot.) If we ask this electorate which candidate they prefer between these two candidates, the electorate chooses Manigo. Thus, we could argue that Manigo should be the election winner.

If we perform the same head-to-head analysis between Manigo and Resnick, we see that 11753 voters prefer Manigo to Resnick while 12352 voters prefer Resnick to Manigo. Since Resnick beats Manigo head-to-head and Manigo beats Hutchinson, it seems by transitivity that Resnick is the strongest candidate and should be the election winner. However, if we analyze the final head-to-head matchup between Resnick and Hutchinson, we see that Hutchinson wins this matchup 12421 to 12165. Thus, in terms of head-to-head matchups this election contains the \emph{Condorcet cycle} Hutchinson beats Resnick who beats Manigo who beats Hutchinson. The aggregate preferences of this electorate are not transitive, a paradoxical outcome. Ideally (at least in social choice theory), the winner of an election is a candidate who defeats all other candidates in head-to-head matchups; such a candidate is called a \emph{Condorcet winner}. This election does not contain a Condorcet winner, and thus it is difficult to decide which candidate is the most deserving of a win.

Elections without Condorcet winners are rarely observed in RCV elections (\cite{GM2}, \cite{MM2}, \cite{S}). Out of hundreds of elections previously checked there is only one previously documented example, a 2021 city council race in Minneapolis \cite{MM1}. The example from Oakland is actually ``cleaner'' than the Minneapolis example because the Minneapolis city council election contained five candidates but voters were allowed to rank only their top three (i.e., voters could not express a complete ranking of the candidates). If voters were allowed to rank all five candidates, it is possible that the Condorcet cycle in the Minneapolis election would disappear. The Oakland School Director election is the first documented election without a Condorcet winner in which voters could provide a complete ranking of the candidates.

Whenever an RCV election does not contain a Condorcet winner or the election contains a Condorcet winner but that candidate is not the election winner, the election will also demonstrate the \emph{spoiler effect}. In the popular discourse around RCV, the spoiler effect is said to occur when the removal of a losing candidate (or, depending on the source, a set of losing candidates) causes the winner to change. Note in the Oakland School Director election if the losing candidate Resnick were removed from the election then Manigo would win instead of Hutchinson, and thus this election demonstrates the spoiler effect. Like Condorcet cycles among the top candidates, the spoiler effect is rarely observed in real-world RCV elections. Out of hundreds of elections checked (\cite{GM2}, \cite{MW}) there have only been three previously documented examples: the 2009 mayoral election in Burlington, VT (\cite{GHS}, \cite{Mi}), the aforementioned city council race in Minneapolis, and the August 2022 Special US House election in Alaska \cite{GM}.

\section{Monotonicity Paradoxes and Compromise Voting}
The Oakland School Director election demonstrates further paradoxical behavior. Suppose that 40 voters who cast the ballot Resnick$\succ$Manigo$\succ$Hutchinson were to shift Resnick down one ranking and cast the ballot Manigo$\succ$Resnick$\succ$Hutchinson instead. What effect should this have on the election outcome? On the one hand, since we have given Manigo more voter support it perhaps makes sense if she were to win the resulting election after giving her 40 more first-place votes, replacing Hutchinson as the winner. On the other hand, we did not change very many ballots so it would also be reasonable for Hutchinson to remain the winner of the resulting election. However, the reader can check that Resnick would win the resulting election if these 40 voters changed their minds in this fashion. That is, losing some voter support would turn the losing candidate Resnick into the winner.

How is such a counterintuitive result possible? Recall that Hutchinson's and Manigo's initial vote totals are 8227 and 8190, respectively, and thus only 37 votes separate these candidates in the first round. If Manigo picks up an additional 40 first-place votes then Hutchinson would be eliminated in the first round and Manigo would advance to face Resnick. Even with the support of these extra 40 voters Manigo does not have enough votes to defeat Resnick in the final round. In fact, the preference profile shows that Resnick could be shifted down one ranking on 598 ballots of the form Resnick$\succ$Manigo$\succ$Hutchinson and he would still have enough votes to defeat Manigo head-to-head. Such a counterintuitive result is possible because we can engineer a change in the order in which candidates are eliminated in such a way that the winner changes in a paradoxical manner.

If an RCV election contains a set of ballots $\mathcal{B}$ and a losing candidate $L$ such that shifting $L$ down the rankings on the ballots in $\mathcal{B}$ creates a new preference profile in which $L$ is the winner, we say that the election exhibits a \emph{downward monotonicity paradox}. Such paradoxes are rarely observed in real-world elections. There have only been two prior documented elections demonstrating this paradox in single-winner RCV elections: a 2021 city council race in Minneapolis \cite{MM1} and a 2020 Board of Supervisors race in San Francisco \cite{GM2}. In \cite{GM2} the authors check 182 RCV elections which go to at least a second round for downward monotonicity paradoxes, and only these two elections are found.

This election also exhibits an \emph{upward monotonicity paradox}, which is like a mirror image of a downward paradox. Note that if 2000 voters who cast the ballot Resnick$\succ$Hutchinson$\succ$Manigo were to shift Hutchinson up one ranking and cast the ballot Hutchinson$\succ$Resnick$\succ$Manigo instead, then Hutchinson would lose the resulting election. To see why, note that in the resulting modified election we have taken enough first-place votes away from Resnick that he is now eliminated first, and Manigo goes on to defeat Hutchinson head-to-head in the final round. Gaining extra support from these 2000 voters would actually cost Hutchinson the election.

If an RCV election contains a set of ballots $\mathcal{B}$ such that shifting the winning candidate $W$ up the rankings on the ballots in $\mathcal{B}$ creates a new preference profile in which $W$ is no longer the winner, we say that the election exhibits an \emph{upward monotonicity paradox}. As with downward paradoxes, upward monotonicity paradoxes are rarely observed in real-world elections. There have only been three prior documented elections demonstrating this paradox in single-winner American RCV elections: the 2009 mayoral election in Burlington, VT (\cite{GHS}, \cite{Mi}); the previously mentioned 2021 city council race in Minneapolis \cite{MM1}; and the August 2022 Special House election in Alaska \cite{GM}. In \cite{GM2} the authors check 182 RCV elections which go to at least a second round for upward monotonicity paradoxes, and only these three elections are found.

This election also has the interesting feature that there exists a set of voters who, by ranking their favorite candidate in first place, caused their least favorite candidate to win. If 1800 of the voters who cast the ballot Resnick$\succ$Manigo$\succ$Hutchinson (and we assume there exist 1800 voters who cast this ballot sincerely) were to instead cast the ballot Manigo$\succ$Resnick$\succ$Hutchinson, then Resnick would be eliminated in the first round and Manigo would defeat Hutchinson 11370 votes to 11322 in the final round. That is, these 1800 voters would have been better off casting a ``compromise vote'' for Manigo, in which case their second-favorite candidate wins instead of their least-favorite candidate. 

If an election contains a set of like-minded voters who could have created a more preferable electoral outcome by not ranking their favorite candidate in first place, we say that the election exhibits a \emph{compromise vote failure}, a notion of strategic voting first rigorously explored in  \cite{GA}. While such an outcome is not paradoxical (at least, it is not as counterintuitive as a monotonicity paradox), it is undesirable if voters have to worry that ranking their favorite candidate in first place creates a worse electoral outcome for them. In \cite{GM2} the authors check 182 RCV elections which go to at least a second round for a compromise vote failure, and only seven elections are found.

\subsection{What about the case when  we process ballots incorrectly?} If we use the ``original'' vote data in which ballots were processed incorrectly as described in Section \ref{sec:ballots}, what happens to the paradoxes exhibited by the election? Recall that in this case, the first-place vote totals for Hutchinson, Manigo, and Resnick were 8112, 8153, and 9954, respectively. Hutchinson was eliminated and Resnick defeated Manigo 12352 to 11753 in the final round\footnote{Note that the mishandled ballots were not discarded from the election, they were simply ignored in the first round of counting. This is why we obtain the same vote totals here as we did in Section \ref{sec:Cond} when comparing Manigo to Resnick head-to-head.}.

In this modified election, a downward paradox is no longer possible. If we shift Hutchinson down on any ballots then he is still eliminated first, and thus we cannot turn Hutchinson into a winner by weakening his voter support. Similarly, if we move Manigo down the rankings on more than 41 ballots on which she is ranked first, she would be eliminated first and we cannot turn her into a winner. Both Manigo and Hutchinson are so close to each other and also so far from the winner Resnick that we cannot engineer a downward paradox in favor of either of them. Thus, mishandling the ballots perhaps produced vote data that is more palatable from a social choice perspective.

However, the modified vote data with mishandled ballots still exhibits an upward monotonicity paradox and also demonstrates a \emph{no-show paradox}, a paradox related to non-monotonicity. To demonstrate the upward paradox, note that if we were to shift Resnick up one ranking on 42 ballots of the form Manigo$\succ$Resnick$\succ$Hutchinson then Manigo would be eliminated first and Resnick would lose to Hutchinson in the final round. The entire kerfuffle over choosing the wrong winner could have been avoided if Resnick had done a better job reaching out to Manigo voters, thereby losing the election even when the ballots were processed incorrectly. 

Furthermore, note that if 42 voters who cast the ballot Manigo$\succ$Hutchinson$\succ$Resnick were to abstain from the election then (assuming the ballots are processed incorrectly) Manigo would have been eliminated first. As a result, Hutchinson would advance to the final round and,  even with these ballots removed from the election, he would still have enough votes to defeat Resnick. This is an example a no-show paradox, where there exists a set a voters who could have created a more desirable electoral outcome by not casting a vote. In the case where the ballots are mishandled, these voters could have gotten their second-favorite candidate as the winner instead of their least-favorite, if they had abstained from the election. 

Thus, in moving from the ``original'' election in which ballots were mishandled to the real election, we swap a no-show paradox for a downward monotonicity paradox. There has never been a documented single-winner RCV election which simultaneously demonstrates downward and upward monotonicity paradoxes as well as a no-show paradox, but this Oakland election seems to have come the closest to doing so.

\section{Different Winners under Different Voting Methods}

The Oakland School Director election is extremely close, as evidenced by the Condorcet cycle among the three candidates. Because of this cycle, each of the three candidates has a reasonable argument to be crowned the winner of the election. Another way to see the closeness of this election is to show that we obtain different election winners when using other classical voting methods from social choice theory. For example, the method of \emph{plurality runoff} works as follows: eliminate all candidates except the two with the most first-place votes, and the winner of the election is the winner of the head-to-head matchup among the two remaining candidates. In the Oakland election the plurality runoff winner is Resnick when we use the real vote data, because when we include write-in candidates the first round vote totals are 8147, 8176, 9977, and 269 for Hutchinson, Manigo, Resnick, and write-ins, respectively. The write-in candidates and Hutchinson are eliminated in the first round, and Resnick defeats Manigo head-to-head in the runoff round. It is rare for the RCV and plurality runoff winners to be different in real-world elections  \cite{MM2}, and this Oakland election is the only ranked choice example of which I'm aware where the difference in the two winners is caused by the presence of a handful of ballots for write-in candidates.

We also consider two other classical voting methods: \emph{plurality} and \emph{Borda count}. Under plurality, the candidate with the most first-place votes wins, and thus Resnick is the plurality winner of the Oakland School Director election. Because he would win under both plurality and plurality runoff, Resnick seems like he has a strong case to be the election winner, given the underlying voter preference data.

The method of Borda count is a points-based method in which the candidate with the highest number of points wins. The traditional definition of the Borda count method assumes that each voter provides a complete preference ranking of the candidates. In this case, if there are $n$ candidates in then a first-place vote for a candidate is worth $n-1$ points, a second-place vote for a candidate is worth $n-2$ points, and so on, until a last-place vote is worth zero points. The candidate with the most points after tallying points from all the ballots wins. When voters cast ballots with incomplete preferences, we must decide how to adapt the points system to such ballots. There are many possible ways to do so; we choose two reasonable approaches. Following \cite{B}, we interpret partial ballots under the \emph{pessimistic model} or the \emph{optimistic model}. Suppose a voter ranks $k<n$ candidates on their ballot. In the pessimistic model,  the top ranked candidate on the ballot receives $n-1$ points, the second ranked candidates receives $n-2$ points, and so on, and any candidate left off the ballot receives 0 points. In the optimistic model, the candidates left off the ballot each receive $n-k-1$ points, so that the point gap between the last ranked candidate and candidates left off the ballot is only 1 point. For example, in a five-candidate election with candidate set $\{A, B, C, D, E\}$, if a voter casts the ballot $A\succ B \succ- \succ - \succ - $ then under the pessimistic model this voter gives 5 points to $A$, 4 points to $B$, and no points to the other candidates. Under the optimistic model, 3 points are given each to $C$, $D$, and $E$.

Table \ref{Borda_table} gives the Borda scores for each candidate under each of the models, both using the data from Table \ref{preference_profile} and the full set of ballots including write-in candidates. Note that the choice of how we process partial ballots matters: Hutchinson wins under the optimistic model when write-ins are disregarded, while Resnick wins under the pessimistic model. From just the three-candidate perspective, there is no single Borda winner. However, if we include write-in candidates in the calculations then Hutchinson actually wins under both models. The reason is that voters in this election were allowed to rank two write-ins on the ballot, meaning that voters could rank five candidates on their ballots. Thus, when write-ins are included a candidate can receive a maximum of four (rather than two) points from a voter and this difference, combined with how write-ins are sprinkled throughout the ballots, allows Hutchinson to narrowly defeat Resnick under either model. As with plurality runoff, we see that this election is so close that a tiny number of write-in votes can affect the election winner.

 \begin{table}[]
  \centering
\begin{tabular}{l|c|c|c}
 & Hutchinson & Manigo &Resnick \\
\hline
Borda Score, OM (Table 1) & 29,329 & 29,190 & 28,690\\
Borda Score, PM (Table 1)& 23,743  & 23,123 & 24,517\\
\hline

Borda Score, OM (full data) & 82,962 & 82,823 & 82,287\\

Borda Score, PM (full data) & 61,969 & 60,831 & 61,480\\

\end{tabular}

\caption{Borda scores under the optimistic model (OM) and the pessimistic model (PM). The scores in the top two rows are calculated using Table \ref{preference_profile}; the scores in the bottom two rows are calculated using the full ballot data including write-ins.}
  \label{Borda_table}
\end{table}

Because Hutchinson is the Borda winner in multiple scenarios and is also the RCV winner, he can make a strong argument that he should win the election given the preference ballot data. Since Manigo defeats Hutchinson in a head-to-head matchup, she can also make a strong argument for herself. Thus, this Oakland School Director election has similar dynamics to the famous 2009 mayoral election in Burlington, VT, in which the plurality, RCV, and Borda winners were all different\footnote{In the Burlington election, the Borda winner did not change under different models of handling partial ballots.}. In an extremely close election, it's almost as if we are choosing a voting method instead of choosing a winning candidate.

\section{Conclusion}

The 2022 District 4 election for School Director in Oakland is the fourth documented RCV election which simultaneously exhibits several of RCV's deficiencies as a voting method. For detractors of RCV, this election perhaps provides another demonstration of why RCV should not be used. It is a reasonable position that any method susceptible to monotonicity paradoxes or the spoiler effect should not be used, and this election provides another example that such issues can arise in practice. Given the discussion about ballot processing in Section \ref{sec:ballots}, this election could also be used to justify using a voting method which does not use preference ballots, such as approval or STAR voting.

While I do not consider myself an RCV advocate, for several reasons I do not think this case study provides a strong argument against the use of RCV. First, the preference ballot data demonstrates a Condorcet cycle among the top candidates and this cycle occurs irrespective of the voting method used. Given such a conflicted electorate, it is not clear that any voting method can choose the ``correct'' or ``most deserving'' candidate, as such a candidate may not meaningfully exist. Monotonicity paradoxes are undesirable, but in this case they seem to occur because the election is so close. The only other documented single-winner election which demonstrates both an upward and downward monotonicity paradox is the previously mentioned city council race in Minneapolis, which also contained a Condorcet cycle among the top three candidates.

Second, while issues such as monotonicity paradoxes are undesirable, they seem to occur rarely in real-world elections. If a Condorcet winner exists then these paradoxes have occurred in only two single-winner American elections, the previously mentioned elections in Burlington and Alaska. There is seemingly no perfect voting method, and thus if a voting method's weaknesses are rarely observed then that seems like the best outcome we can hope for.

Third, once the ballot processing issue was sorted out RCV chose Hutchinson as the winner, and he seems to be the best choice given the preference data. Hutchinson wins under both models of partial ballots when using Borda count (if we include the complete data with write-ins), and he wins under the optimistic model when disregarding write-in candidates (the optimistic model seems to be the ``best'' way to adapt partial ballots to Borda count; see \cite{K}). Hutchinson also receives the highest score if we add up the number of first and second place preferences for each candidate (i.e., he is the winner under Bucklin voting). Also, all three candidates lose one head-to-head matchup, and Hutchinson's loss margin in his loss is smaller than the loss margins for the other two candidates in their losses. In this sense, Hutchinson is the ``closest'' candidate to being a Condorcet winner. If RCV is able to select the ``correct'' winner given the confused preferences of this electorate then perhaps the election's social choice issues become less salient.

\section*{Acknowledgements}

Thank you to Jeanne Clelland for  telling me there was a ballot processing error in this election.

\end{document}